\documentclass[10pt]{iopart}

\usepackage{iopams}		
\usepackage{graphicx}   
\usepackage{epstopdf}	
\usepackage{verbatim}   
\usepackage{color}      
\usepackage{subfigure}  
\usepackage{hyperref}   
\usepackage{braket}

\begin{document}

\title{Nb/InAs nanowire proximity junctions from Josephson to quantum dot regimes}

\author{Kaveh Gharavi$^{1,2}$, Gregory W. Holloway$^{1,2,3}$, Ray R. LaPierre$^4$, and Jonathan Baugh$^{1,2,3,5}$}
\address{$^1$ Institute for Quantum Computing, University of Waterloo, Waterloo, Ontario N2L 3G1, Canada}
\address{$^2$ Department of Physics and Astronomy, University of Waterloo, Waterloo, Ontario N2L 3G1, Canada}
\address{$^3$ Waterloo Institute for Nanotechnology, University of Waterloo, Waterloo, Ontario N2L 3G1, Canada}
\address{$^4$ Department of Engineering Physics, Centre for Emerging Device Technologies, McMaster University, Hamilton, Ontario L8S 4L7, Canada}
\address{$^5$ Department of Chemistry, University of Waterloo, Waterloo, Ontario N2L 3G1, Canada}
\ead{baugh@uwaterloo.ca}

\date{\today}

\begin{abstract}

The superconducting proximity effect is probed experimentally in Josephson junctions fabricated with InAs nanowires contacted by Nb leads. Contact transparencies $t \sim 0.7$ are observed. The electronic phase coherence length at low temperatures exceeds the channel length. However, the elastic scattering length is a few times shorter than the channel length. Electrical measurements reveal two regimes of quantum transport: (i) the Josephson regime, characterized by a dissipationless current up to $\sim 100$ nA, and (ii) the quantum dot regime, characterized by the formation of Andreev Bound States (ABS) associated with spontaneous quantum dots inside the nanowire channel. In regime (i), the behaviour of the critical current $I_c$ versus an axial magnetic field $B_{||}$ shows an unexpected modulation and persistence to fields $>2$ T. In the quantum dot regime, the ABS are modelled as the current-biased solutions of an Anderson-type model. The applicability of devices in both transport regimes to Majorana fermion experiments is discussed. 
\end{abstract}

                              
\submitto{\NT}
\maketitle
\ioptwocol


\section{\label{sec:intro}Introduction}

Andreev reflection at an interface between a superconductor and a normal metal (or semiconductor) leads to a number of surprising transport phenomena, including a subharmonic gap structure due to multiple Andreev reflection (MAR) \cite{flensberg_1988_otbk_corrected}, Andreev Bound States (ABS) \cite{kulik_1969, Pillet_NatPhys2010}, and supercurrent in superconductor-normal-superconductor (SNS) junctions \cite{Klapwijk2004,Doh2005}. At low temperature, the normal section length that can support a supercurrent is limited by the phase coherence length of conduction electrons in that section, and in semiconductors this can far exceed the superconducting coherence length that characterises the superconductor. While planar SNS junctions are well studied \cite{Barone1982} and have revealed interesting phenomena such as Fraunhofer interference in a magnetic field \cite{chiodi_PRB2012}, there are fewer studies on semiconductor nanowire-based SNS junctions. The latter are especially relevant in light of recent advances in the search for Majorana fermions, zero-energy quasiparticles at the boundaries of one-dimensional topological superconductors \cite{kitaev_2001,Alicea2011}. There are several reported observations of signatures of Majorana fermions based on proposals for their detection in semiconducting nanowires contacted with superconductors \cite{Albrecht2016, Zhang_Delft2016_MF, Mourik2012, Das2012,Deng_nanolett2012}. The semiconductor must have a large spin-orbit coupling and make high transparency contacts with a superconductor, properties shared by InAs and InSb nanowires. While Al/InAs junctions are relatively well studied \cite{Jellinggaard_PRB2016,Albrecht2016,ChangW_nnano2015}, Nb/InAs is less well characterized, but potentially advantageous due to a significantly larger superconducting gap $\Delta$ and that Nb is a type II superconductor, with an upper critical field $H_{c2}\sim 2.8$ T. Given that nanowire junctions are typically in the diffusive transport regime, a key question relevant to Majorana and other research is how (static) disorder in the nanowire is manifested in quantum transport experiments on SNS devices. 

Here, we report on proximity effect Josephson junctions made with InAs nanowires contacted by Nb leads. The junctions have channel lengths $L \sim 200$ nm, shorter than the estimated electronic phase coherence length $\xi \sim 250 - 350$ nm at low temperature, but longer than the estimated elastic scattering length by a factor of $3 - 5$. Data is presented for 5 different devices (d1 -- d5) falling within two distinctive regimes of quantum transport: devices with higher electron mobilities $\mu \sim 18,000$ cm$^2$/(Vs) are characterized by the observation of a disspationless current (`Josephson regime'), whereas devices with lower mobilities $\mu \lesssim 10,000$ cm$^2$/(Vs) typically show signatures of spontaneous, unintentional quantum dots (`quantum dot regime') and signatures of induced superconductivity in the form of Andreev Bound States (ABS).

In the Josephson regime, the diffusive nature of the nanowire transport is manifested by a critical current that is modulated as the nanowire chemical potential is varied, closely following the normal state conductance fluctuations, but with enhanced sensitivity. Transmission coefficients in the OTBK model \cite{flensberg_1988_otbk_corrected} are found to be in the range $\sim 0.4 - 0.7$, indicating sufficiently good contact transparencies to observe Andreev physics. The junction phase dynamics are typically overdamped. In devices with more disorder in the potential landscape, the quantum dot (QD) regime is observed and reveals resonances associated with ABS up to temperatures $\sim 2$ K. Unlike previously reported observations \cite{Jellinggaard_PRB2016, ChangW_nnano2015, ChangW_PRL2013, SandJespersen_PRL2007,DeFranceschi_nnano2014,Kumar_PRB2014_CNT_ABS, Gaass_PRB2014_CNTABS, Pillet_PRB2013, Kim_PRL2013_CNTABS, Pillet_NatPhys2010, eichler_PRL2007_CNTABS}, the gate dependence of the ABS resonances does not appear to depend strongly on whether the QD is in a even- or odd-electron state. In order to explain this, we adapt an Anderson-type mode \cite{kirsanskas_ysr_2015} originally developed for a phase-biased S-QD-S configuration to our experiment, which behaves as a current-biased S-N-QD-N-S configuration. The model qualitatively explains how the observed patterns of ABS resonances depend on a competition between the strength of the coupling to superconductivity and the dot addition energy $E_{add}$. The presence of ABS associated with either spontaneous or engineered QDs could be useful as energy filters in transport spectroscopy of zero modes. 

In section \ref{sec:expt}, we discuss fabrication methods and basic device characteristics. Section \ref{sec:results} presents the low temperature transport results in the Josephson and QD regimes, and introduces a model for the ABS associated with QDs. In section \ref{sec:discussion}, we discuss these results in the context of related observations in the literature, and with regard to the relevance of these types of devices in exploring Majorana fermion physics. Conclusions are given in section \ref{sec:conclusions}. 

\section{\label{sec:expt}Experiment}

\begin{table*}[t!]
\caption{Basic device characteristics. $L$ is the channel length after lift-off of contacts, typically $\sim 30$ nm shorter than the channel length written on PMMA from the lithography stage. The field-effect mobility $\mu$ (measured at $\sim 50$ K) can serve as a predictor as to whether a junction will be in the Josephson or the QD regime, with $\mu \lesssim 1.0$ m$^2$/(Vs) typical of the QD regime. \label{tab:device_summary}}
\begin{tabular}{@{}llllllll}
\br
		& Nanowire  			& Ar ion mill	 & Contact           	 & Gate            & L      	& $\mu$       	& Josephson /  \\
	   & type      				& pressure		 & Recipe            	 & Layout          & (nm)   	& m$^2$/(Vs) 	& QD regime    \\ \mr
d1     & Al$_2$O$_3$/InAs   	& 0.2 Pa		 & Ti/Nb (2 nm/80 nm)	 & Global backgate & 170 nm 	& $1.8 \pm 0.3$ & Josephson    \\
d2     & InAs      				& 0.6 Pa		 & Nb (50 nm)        	 & 5 local gates   & 200 nm 	& $1.6 \pm 0.3$	& Josephson    \\
d3     & Al$_2$O$_3$/InAs   	& 0.6 Pa		 & Ti/Nb (2 nm/80 nm)	 & Global backgate & 170 nm 	& $0.8 \pm 0.2$	& QD           \\
d4     & InAs      				& 1.3 Pa		 & Ti/Nb (2 nm/50 nm)	 & Global backgate & 200 nm 	& $0.6 \pm 0.2$	& QD           \\
d5     & Al$_2$O$_3$/InAs   	& 0.6 Pa		 & Ti/Nb (2 nm/80 nm)	 & Global backgate & 170 nm 	& $0.9 \pm 0.2$	& QD           \\ \br  
\end{tabular}
\end{table*}

The devices d1 -- d5 reported on here are based on a single batch of undoped InAs nanowires grown in a gas-source molecular beam epitaxy system, starting from a gold nanoparticle. The diameter of the selected nanowires is in the range $40 - 65$ nm. Details of the nanowire growth can be found in Refs. \cite{Gupta_2013_mobility, Holloway_2013_CoreShell}. Devices d3 and d5 were on the same chip; otherwise each device is representative of a distinct fabrication run. Thirty-six devices were measured in total; all of the devices not reported on explicitly here were measured at a temperature of $1.5$ K and showed characteristics similar to the ones reported. While none of these devices show a supercurrent at 1.5 K, we estimate from the minimum resistances that $\sim 1/4$ of the devices would show a supercurrent in the range of $10 - 100$ nA at dilution refrigerator temperatures. 

For devices d1, d3 and d5, an $8$ nm thick Al$_2$O$_3$ shell was deposited via atomic layer deposition onto the nanowires on their growth substrate, covering all facets of the nanowires. Our previous studies have found \cite{Holloway_2016_odt} that the desirable transport characteristics of these nanowires -- such as mobility, low gate hysteresis, stability at low temperature, etc. -- can be maintained with an optimized process of chemical surface passivation (using octadecanethiol \cite{Sun_nanolett2012_odt_passivation, Hang_nanolett2008_odt_passivation, Petrovykh_2009_odt}) followed by a conformal Al$_2$O$_3$ shell deposition. These coated nanowires, as well as bare nanowires, are moved to n++ Si / 300 nm SiO$_2$ device substrates via dry deposition. Before nanowire deposition, preparatory fabrication steps are performed, such as placing Ti/Au bonding pads and alignment markers. Other steps varied across the devices studied. Device d2 has a set of five evenly spaced bottom gates in the channel region, covered by a $20$ nm SiN$_x$ dielectric layer. Device d4 is an entrenched nanowire, where reactive ion etching was used to etch a $\sim 60$ nm wide and deep trench in SiO$_2$ in which the nanowire was positioned. The n++ Si substrate serves as a global backgate for all devices except d2. Superconducting contacts are defined using electron beam lithography on a bilayer PMMA resist of thickness $\sim 500$ nm. Following a short $5 -12$ s HF etch to remove the Al$_2$O$_3$ shell and/or the native oxide of InAs, the device substrate is quickly moved to the deposition chamber. In-situ Ar ion milling (rf reverse sputtering) is performed at Ar pressures listed in table \ref{tab:device_summary} using an rf power $50$ W for a duration $\sim 10$ minutes, followed by deposition of Nb or Ti/Nb contacts via dc magnetron sputtering. Table \ref{tab:device_summary} summarizes the contacting recipes for the 5 devices, as well as other junction properties. After lift-off of the contacts, transport measurements are carried out at low temperature in one of two cryostats: a dilution refrigerator with a base lattice temperature $T_{\mathrm{L}} = 25$ mK (for all Josephson and some QD devices), and a pumped $^4$He system with base $T_{\mathrm{L}} = 1.5$ K (for QD devices). The electron temperatures are $\sim 100$ mK and 1.5 K, respectively. Electrical measurements are made by applying a dc bias voltage (or current) using a custom high-resolution source, and the dc voltage and current responses of the device measured with low-noise preamplifiers.

Table \ref{tab:device_summary} reports the field effect mobilities $\mu$ obtained from transconductance measurements at $T \simeq 50$ K via the analysis given in Ref. \cite{Gupta_2013_mobility}. The capacitance between the nanowire and the global backgate is estimated using a COMSOL model. The model shows the effect of the metallic leads of the short channel junction is to screen the electric field of the gate, reducing the effective capacitance by a factor $4 - 5$ for channel lengths $L \sim 200$ nm. Uncertainty in this screening factor is the main source of uncertainty in the values of $\mu$ reported in table \ref{tab:device_summary}.

The field effect mobility $\mu$ can predict whether a particular device will exhibit a large supercurrent (i.e. the Josephson regime) at base temperature. Most devices display $\mu \leq 10^4$ cm$^2 /$(V.s), corresponding to an elastic mean free path $l_e \lesssim 40$ nm, and show signatures of unintentional quantum dots forming inside the nanowire channel (i.e. the QD regime). Such devices have not been observed to sustain a supercurrent larger than $2$ nA at $T_{\mathrm{L}} = 25$ mK. However, devices such as d1 and d2 with $\mu \gtrsim 1.5 \times 10^4$ cm$^2 /$(V.s), i.e. $l_e \gtrsim 60$ nm, typically do not show Coulomb diamonds in the low temperature I-V data. Given channel lengths $L = 170 - 200$ nm, these devices are in the diffusive regime, but a lack of Coulomb blockade allows for resistances as low as $3\; \mathrm{k} \Omega$ at $V_g \simeq 10$ V. These Josephson regime junctions carry supercurrents up to $\sim 100$ nA at $T_{\mathrm{L}} = 25$ mK. We expect that optimizing the InAs surface preparation recipe prior to the sputtering of the Nb contacts and improving the Nb/InAs contact transparency should increase the likelihood for a device to exhibit Josephson behavior. No signatures of proximity superconductivity were observed in devices fabricated without Ar ion milling. Lower Ar ion milling pressure (such as for d1) has been found to correlate with reduced Nb/InAs contact resistance; however, one must also consider the natural variation between different nanowires when interpreting this observation, since a total sample size of $\sim 10$ devices per fabricated chip is relatively small. Even for devices with high Nb/InAs contact transparency, it is the potential landscape in the nanowire channel that plays a dominant role in determining which regime of quantum transport is observed. The salient features of these two regimes are presented in the next section.

\section{\label{sec:results}Results}

\subsection{Josephson regime}

\begin{figure}[t!]
	\centering
	\includegraphics[width=8.5cm]{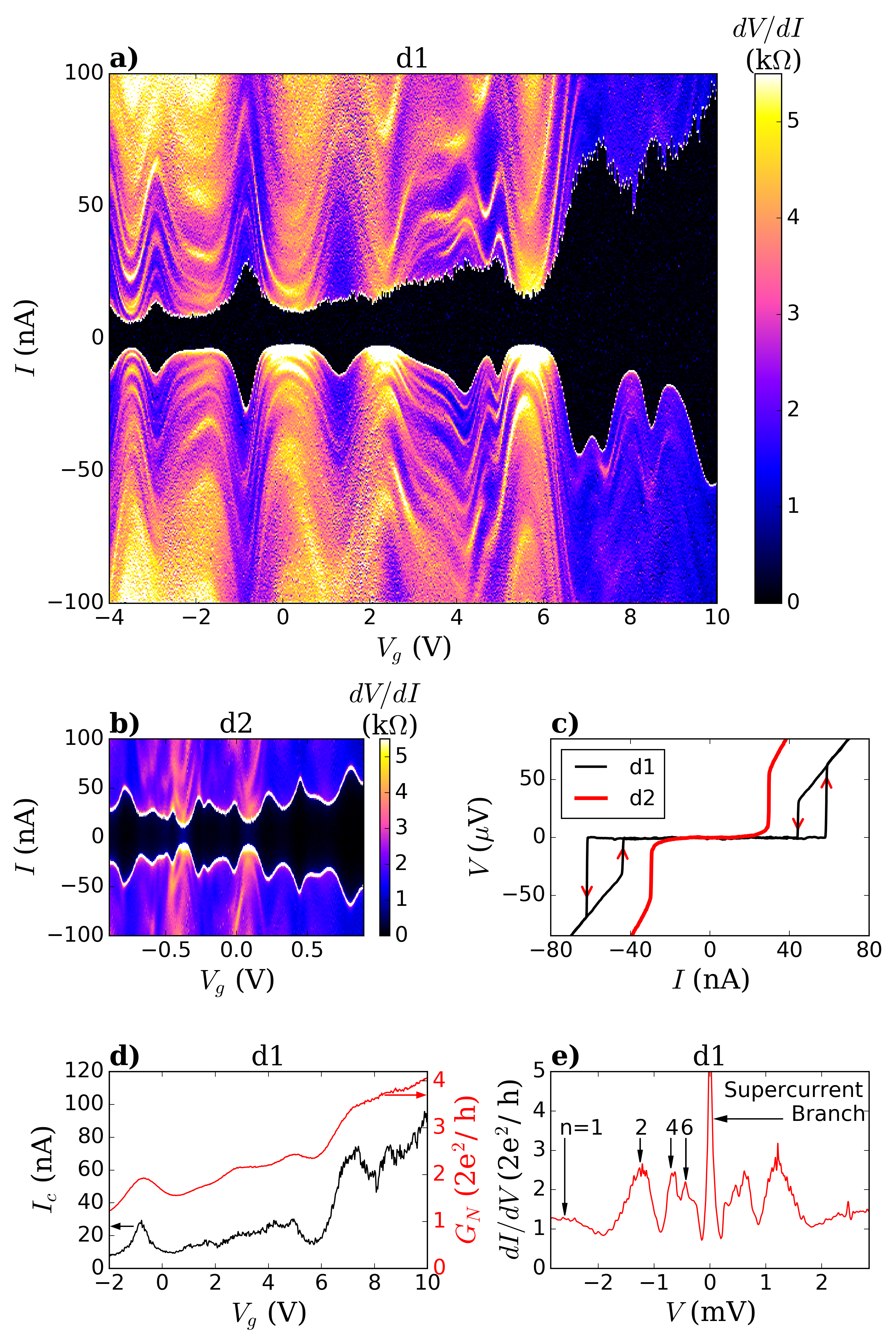}
	\caption{Josephson regime devices d1 and d2. \textbf{a,b)} Differential resistance $dV/dI$ versus bias current $I$ and gate voltage $V_g$ for d1 and d2, respectively. The supercurrent magnitude is modulated by the gate voltage. \textbf{c)} Typical $I$-$V$ traces for d1 (black) and d2 (blue/dark grey). The residual resistance observed for d2, compared to sharp normal to dissipationless transitions for d1, suggest that the Nb/InAs interface may be of higher quality in d1. \textbf{d)} A strong correlation is observed between the critical current $I_c$ and the normal state resistance $G_N$ versus gate voltage $V_g$ (device d1). \textbf{e)} Multiple Andreev reflection (MAR) signatures of order $n$ are observed in d1 as peaks in $dI/dV$, indicated by arrows. It is concluded that the Josephson junction is phase-coherent across the channel. Similar data is observed for d2 (not shown). For all panels, data is measured at $T_\mathrm{L} = 25$ mK, except $G_N$, which is measured at $8$ K.
	  \label{fig:1_ic}}
\end{figure}

We first focus on devices d1 and d2 that are representative of the Josephson regime. Figure \ref{fig:1_ic}a shows the numerically calculated differential resistance $dV/dI$ of d1 versus gate voltage and bias current at a base lattice temperature $T_\mathrm{L} = 25$ mK and zero magnetic field. As the bias current $I$ is swept from negative to positive values, the dc voltage response $V$ of the device is recorded. A dissipationless current (black region) is observed, with a critical current $I_{c}$ whose value can be tuned with the back gate voltage $V_g$, and can assume a value as large as $I_{c} = 97$ nA at gate voltage $V_g = 10$ V. A similar plot is shown in figure \ref{fig:1_ic}b for d2, where all five bottom gates are swept together but $V_g$ refers to the value of the middle gate which was most effective. A maximum critical current of $I_{c} = 55$ nA is observed for d2. Figure \ref{fig:1_ic}c shows the $I$-$V$ traces for both junctions; d1 typically shows a sharp supercurrent transition and a hysteretic behaviour with respect to sweep direction due to quasiparticle heating \cite{Courtois_2008_Joule_heating}, as is typical of nanowire based Josephson junctions. The device d1 was measured at temperatures in the range $T_\mathrm{L} = 25$ mK to $T_\mathrm{L} = 1.2$ K, as well as at $T_\mathrm{L} = 8$ K, inside the dilution refrigerator. The hysteretic behaviour is suppressed at higher temperatures; at $T_\mathrm{L} = 1.2$ K, the retrapping current $I_r$ is reduced by about $20 \%$ (averaged over $V_g$) compared to $T_\mathrm{L} = 25$ mK, but there is negligible difference between the retrapping and critical switching currents. Hence, the local Joule heating that reduces $I_r$ compared to $I_c$, causing hysteresis at $T_\mathrm{L} = 25$ mK, is overshadowed by thermal energy at $T_\mathrm{L} = 1.2$ K. 

The critical temperature of the junction is extrapolated from $I_c$ versus $T$ data to be approximately $T_c = 1.6$ K. On the other hand, d2 shows residual resistance on the order of 100 $\Omega$ near its switching point, and negligibly small hysteresis. This difference in the $I$-$V$ behaviour of the junctions could be due to two effects: higher electron temperature and/or the quality of the Nb/InAs interfaces. First, while d2 was measured at the lattice temperature $T_\mathrm{L} = 25$ mK, we suspect the effective electron temperature might have been higher than $\sim 100$ mK due to high-frequency noise on the five local bottom gates. Secondly, the interface quality can be estimated from the OTBK \cite{flensberg_1988_otbk_corrected} transmission coefficient $t$. The high-bias I-V traces for both devices extrapolate to a finite excess current \cite{Doh2005, flensberg_1988_otbk_corrected} $I_\mathrm{exc} \sim 10 - 100$ nA at zero voltage bias, depending on the gate voltage. This excess current can be used to calculate $t$. In a non-ballistic junction, $t$ can be influenced by scattering processes inside the semiconducting channel. Indeed, the calculated $t$ follows the gate voltage dependence $I_\mathrm{exc}$. However, $t$ can serve as a (conservative) estimate of the sputtered Nb/InAs contact transparency. The obtained value is generally higher when the device is more conductive. For d1, $t$ is within the range $t = 0.56$ to $t = 0.72$ with an uncertainty $\pm 0.02$ at each point within that interval. This is a relatively good value for Nb/InAs devices \cite{Guenel2012,Takayanagi_1995_2D_InAs}. For device d2, lower values are observed, $t = 0.41$ to $t = 0.55$ with uncertainty $\pm 0.02$. 

The phase dynamics of both junctions are overdamped. With both devices tuned to the regime of maximal supercurrent, the quality factor $Q = R_N \sqrt{2eI_{c}C/\hbar}$ is calculated using the resistively and capacitively shunted junction model \cite{Tinkham1996_p202}. Here $R_N$ is normal state resistance and $C$ the source-drain capacitance. For d1, $Q \cong 0.9$. Thus, d1 is close to the transition point between overdamped and underdamped regimes. For d2, $Q \cong 0.1$. This difference is due to a higher critical current in d1 and a larger source-drain capacitance due to thicker sputtered contacts. It should be possible to engineer similar nanowire junctions to be in the underdamped regime by making the Nb contacts with larger cross-sectional area. In a SQUID geometry, this would allow phase dynamics experiments, such as measurement of the current-phase relation \cite{DellaRocca_2007_CPR}, to be performed.

Figure \ref{fig:1_ic}d shows the critical current $I_c$ (black curve) and normal state conductance $G_N$ (red/grey curve) of d1 versus the backgate voltage $V_g$. The $I_c$ curve is extracted from figure \ref{fig:1_ic}a using a threshold resistance: the current bias point at which the numerical $dV/dI$ first goes from $\simeq 0$ (supercurrent branch) to above a threshold resistance $\sim 100\; \Omega$ is identified as the switching point. $G_N$ is measured directly using a standard lock-in technique, with a $2$ nA bias current signal at $17$ Hz and a temperature $T_\mathrm{L} = 8$ K, during the same cooldown. The two quantities show a strong positive correlation, typical of SNS junctions \cite{Delsing_2014_InAs, Guenel2012, Nilsson_2012_InSb, Doh2005, Takayanagi_1995_2D_InAs}. The rms value of universal conductance fluctuations \cite{Jespersen_2009_UCF, Doh2005, Lee_1985_UCF} in $G_N$ versus gate voltage can be calculated by first subtracting a baseline curve from the $G_N$ versus $V_g$ data, then calculating the standard deviation. The baseline is obtained by smoothing the $G_N$ data over intervals $\delta V_g \sim 50$ mV.  We obtain $\delta_{{G_N}} = 4.8 \times 10^{-7}$ S, whereas the fluctuations in $I_c$ display an rms value $3.8$ nA with respect to a similarly obtained baseline. Note that the device is in the overdamped regime, so values extracted for $I_c$ at any gate voltage is reproducible over repeated measurements to within $\pm \sim 0.5$ nA. Comparing the fluctuation ratios of the two quantities shows $\delta_{{G_N}} / G_N = 0.2 \% \ll \delta_{I_c} / Ic = 4 \%$, suggesting that the critical current is a more sensitive probe of mesoscopic fluctuations in the junction compared to normal state conductance, as first noticed in Ref. \cite{Doh2005} for an InAs nanowire junction. The `figure of merit' product $I_c R_N$ has a value in the range $0.2 - 0.4$ mV. Here, $R_N$ is the normal state resistance of the junction. Figure \ref{fig:1_ic}e shows differential conductance $dI/dV$ of d1 versus bias voltage $V$ at $V_g = 0.9$ V. Signatures of multiple Andreev reflection (MAR) are observed at voltages $V_n = 2 \Delta / (e n)$ as peaks in conductance for integer $n$, as indicated by arrows. Here, $e$ is the electronic charge and $\Delta = 1.25$ meV is the superconducting gap of the Nb contacts. It is seen that the product $e I_c R_N$ is factor of $3 - 4$ smaller than $\Delta$. Whereas in an ideal, ballistic SNS junction a value $I_c = \Delta /(e R_N)$ is expected, the observed supercurrent is smaller due to a combination of the diffusive nature of the junction, as well as a residual resistance present at the Nb/InAs interfaces.
 
\begin{figure}[t]
	\centering
	\includegraphics[width=8.5cm]{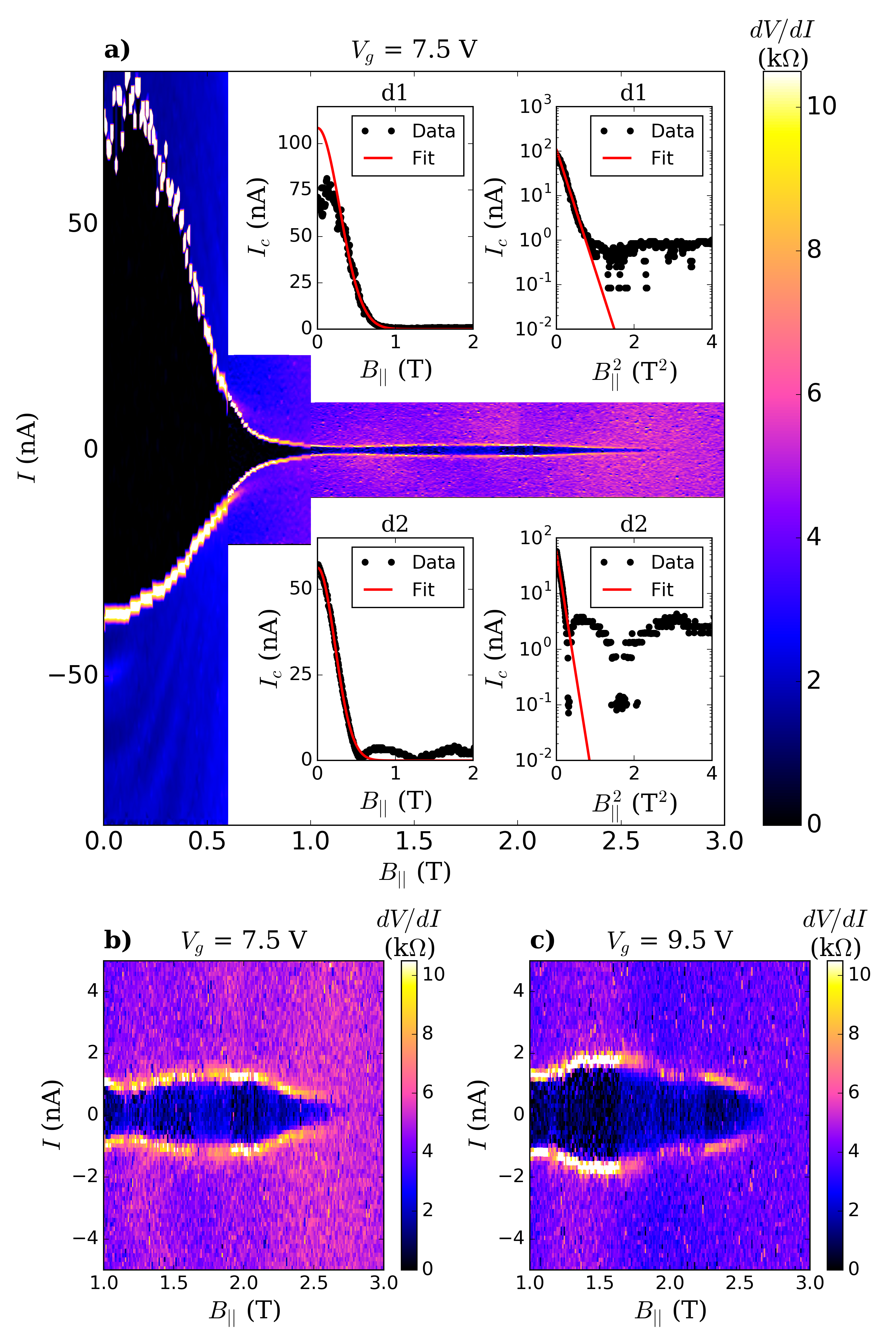}
	\caption{Magnetic field dependence of the critical current $I_c$ for d1 and d2 at $T_\mathrm{L} = 25$ mK. \textbf{a)} Differential resistance $dV/dI$ of d1 versus bias current $I$ and axial magnetic field $B_{||}$ at $V_g = 7.8$ V. Above $0.8$ T, $I_c$ does not follow a Gaussian decay and is weakly modulated with the field, persisting until $\sim 2.8$ T. \textbf{Inset}, Upper: Gaussian fit (red/grey) to extracted $I_c$ vs $B_{||}$ data for d1 on linear and logarithmic scales (left, right, respectively). Depairing mechanisms are accounted for by the Gaussian curve, as discussed in the text. Lower insets: data for device d2 plotted in a similar format as the upper insets. Above $\sim 0.6$ T, d2 shows a stronger oscillation-like modulation with field compared to d1. \textbf{b,c)} High field data ($B_{||} > 1$ T) for two values of $V_g$ show that the $I_c$ variation is not monotonic, and the detailed behaviour depends on gate voltage. 
	\label{fig:2_B-dep}}
\end{figure}

\subsection{Finite magnetic field}

An axial magnetic field $B_{||}$ is a key ingredient predicted by theory to enable tuning the junction into the topological regime \cite{Albrecht2016, Zhang_Delft2016_MF, Mourik2012, Alicea2011}. Signatures of the topological phase transition have also been predicted to appear in the critical current \cite{Aguado_PRB2014_IcSignature}. Although we do not see explicit signatures of topological states here, we focus on the critical current $I_c$, which shows a non-monotonic and complex dependence on $B_{||}$, worthy of further study. 

Figure \ref{fig:2_B-dep}a shows numerical differential resistance $dV / dI$ of d1 versus $B_{||}$ at a backgate voltage $V_g = 7.5$ V. The current is swept from negative to positive values, so the transition at $I > 0$ ($I < 0$) indicates the critical current $I_c$ (the retrapping current $I_r$). Three observations are of note here: (i) In the low-field regime, there is a slight \textit{increase} of $I_c$ with the magnetic field up to $B_{||} \simeq 0.14$ T. This behaviour occurs for a wide range of gate voltages, and can be understood as a manifestation of weak localization due to disorder in the diffusive nanowire channel with $l_e \simeq 60$ nm $< L = 170$ nm. As the magnetic field is increased, back-scattering is suppressed slightly due to the breaking of time-reversal symmetry. This results in an increase of the normal state conductance $G_N$ and therefore $I_c$. However, it is of note that this behaviour is not directly observed in lock-in measurements of $G_N$ vs $B_{||}$ at $T_\mathrm{L} = 8$ K. This is likely explained by temperature: at $T_\mathrm{L} = 25$ mK, the phase coherence length $\xi$ of the nanowire channel of the junction is estimated to be $\xi = 250 - 350$ nm, based on the suppression field $B_s = 0.2 - 0.3$ T above which the weak localization behaviour is no longer observed \cite{Datta_book_p212}. However, at $T_\mathrm{L} = 8$ K, $\xi$ is bounded by the thermal length $L_T = \sqrt{\hbar D / 2 \pi k_B T} \simeq 100$ nm, shorter than the channel length $L = 170$ nm. Thus, the junction cannot be expected to be phase-coherent at $8$ K; indeed no signatures of a supercurrent or MAR are observed there. Similarly, the retrapping current $I_r$ does not show an increase with $B_{||}$, as it is likely bound by the local Joule heating mechanism \cite{Courtois_2008_Joule_heating}. We conclude that, at $T_\mathrm{L} = 25$ mK, the supercurrent can serve as a sensitive probe of the mesoscopic and Fabry-P{\'e}rot type resonances of the junction. (ii) Above $B_{||} \sim 0.3$ T, the magnetic depairing mechanism \cite{Hammer2007} sets in, and the $I_c$ vs $B_{||}$ curve can be fit to a Gaussian, $I_c \propto \mathrm{exp}(-0.526 \Phi^2 / \Phi_0^2)$ (see insets of Figure~\ref{fig:2_B-dep}). Here, $\Phi = \pi d^2 B_{||} / 4$ is the magnetic flux through the axial cross section of the nanowire for diameter $d$, and $\Phi_0 = h/(2e)$ is the superconducting flux quantum. A theoretical calculation \cite{Hammer2007} of the depairing of a planar SNS junction in a perpendicular field predicts $I_c \propto \mathrm{exp}(-0.238 \Phi^2 / \Phi_0^2)$. The reason behind the discrepancy in the numerical prefactor of the exponent between the theory and the experiment is not fully understood. A flux focusing effect is not expected to be at play here, as $B_{||}$ is applied in the device substrate and parallel to the nanowire axis. In fact, the effect of the Nb leads will be to produce a magnetic screening effect, similar to the effect seen in ref. \cite{Strambini_2016}. A full description of the system may require a numerical solution of 3-dimensional Usadel equations \cite{Dubos_2001_long_SNS, Belzig_1999_quasiclassical, Usadel_1970} for the SNS junction while taking into account any screening effects due to the superconducting contacts. These considerations are beyond the scope of this paper. (iii) Interestingly, in the high-field regime it is observed that the Gaussian suppression of $I_c$ does not continue beyond $B_{||} \sim 0.8$ T (see figure \ref{fig:2_B-dep}a insets), and a finite $I_c$ can be resolved up to $B_{||} \sim 2.8$ T. Furthermore, in this high field regime, $I_c$ does not decay monotonically with $B_{||}$, but is modulated with a pattern that depends on the gate voltage. These modulations are not consistent with a Fraunhofer pattern. Figure \ref{fig:2_B-dep}(b, c) shows this high-field behaviour in device d1 for two values of $V_g$. A more oscillatory modulation of $I_c$ vs $B_{||}$ is observed in device d2, following an initial Gaussian decay (bottom inset of figure \ref{fig:2_B-dep}a). The magnitude of $I_c$ modulations in d2 are larger than in d1, with the ratio $I_c(B_{||}) / I_c (0)$ up to $10 \%$ at high fields. Qualitatively similar modulation of $I_c$ vs $B_{||}$ has been theoretically calculated to result from a Josephson interference due to orbital angular momentum states inside an idealized nanowire channel \cite{Gharavi_PRB2015}, with oscillation periods and node positions on the same scale as those in Figure \ref{fig:2_B-dep}. Another possible explanation is an effect due to the Zeeman splitting of the two spin channels in the presence of the axial field and spin-orbit coupling \cite{yokoyama_spin_orbit_zeeman}. However, we estimate the first `node' of oscillation within that theory to occur at $B_{||} \gtrsim 3$ T for an InAs nanowire with Land\'e $g$-factor close to $10$, whereas the first minimum in $I_c$ is observed at $B_{||} \sim 0.8$ T in the experiment. The modulations could also simply be due to the evolution of mesoscopic interference (universal conductance fluctuations due to static disorder) with magnetic field. None of these hypotheses, however, explain the persistence of $I_c$ up to relatively high magnetic field. For InAs nanowires, it is well known that band bending due to surface states enhances the conductivity near the surface. It is possible that a significant fraction of $I_c$ is carried by surface conducting channels that are not necessarily continuous around the perimeter due to the hexagonal faceted geometry. In this case, the effective flux enclosed by those channels could be very small, reducing the response to an axial magnetic field. 

\begin{figure}[t!]
	\centering
	\includegraphics[width=8.5cm]{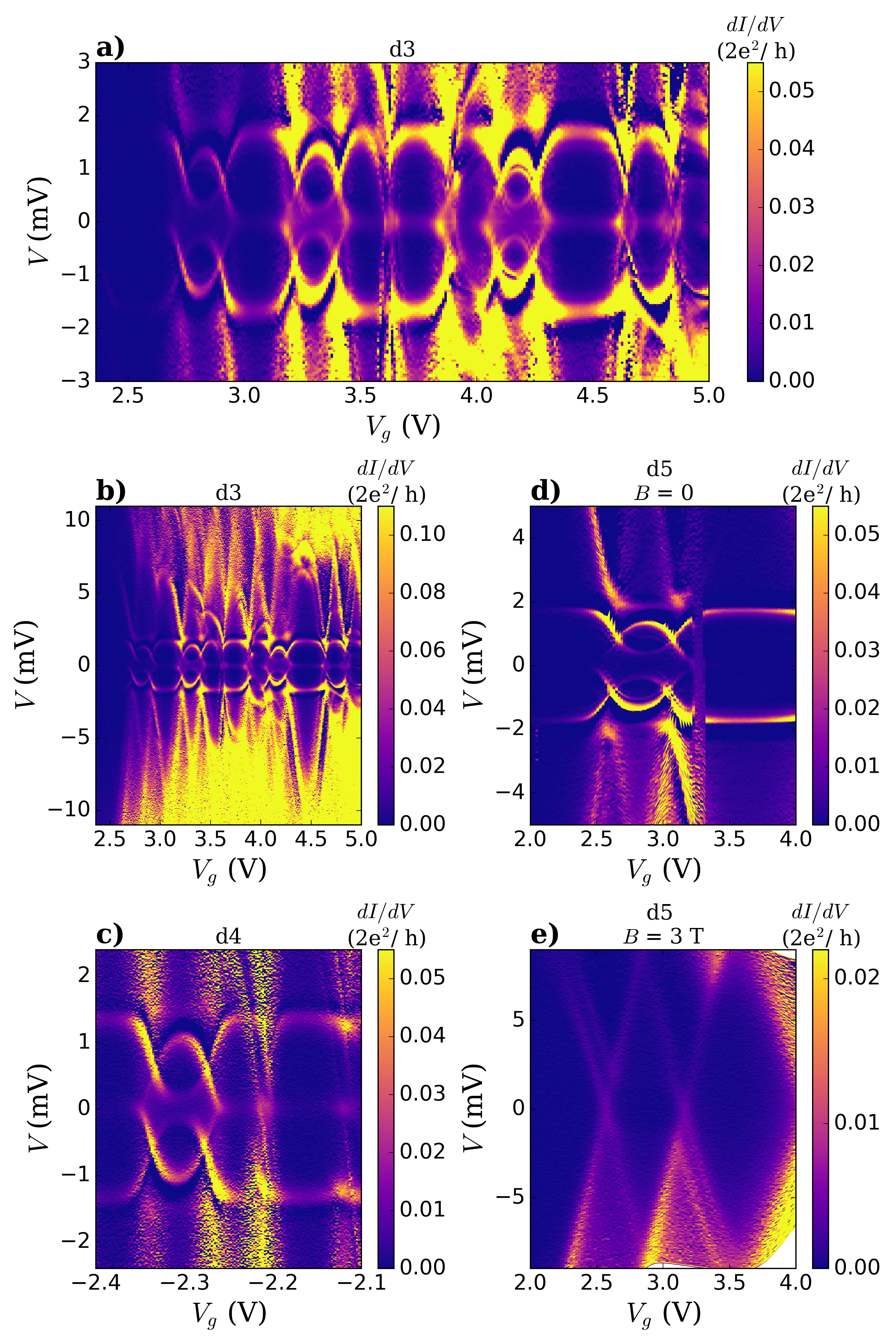}
	\caption{Differential conductance $dI/dV$ of the devices d3--d5 in the QD regime. \textbf{a,b,c)} Zero magnetic field $dI/dV$ data of d3, d4, d5, respectively, versus bias voltage $V$ and gate voltage $V_g$. Coulomb diamonds associated with the spontaneous quantum dots are visible, as well as resonant features identified as Andreev Bound States (ABS) at $|V| < \Delta/e = 1.4$ mV. For each value of $V_g$, four ABS are visible. In Coulomb diamonds with electron addition energy $E_{add} \gtrsim 10$ meV, the inner two ABS merge to form a zero bias peak (ZBP), whereas the outer two ABS are pinned at $|V| = \Delta/e$. By contrast, in Coulomb diamonds with $E_{add} \lesssim 10$ meV, four non-degenerate ABS are visible at $0< |V| < \Delta/e$. \textbf{d)} Same as panel \textbf{c}, but at a magnetic field $B = 3$ T perpendicular to the device substrate, which suppresses the superconductivity in the Nb leads. Comparing panels \textbf{c} and \textbf{d} allows distinguishing the ABS from the Coulomb diamond features, and allows the charge degeneracy points to be identified. For all panels, the data was acquired $T_{\mathrm{L}} = 1.5$ K.
	  \label{fig:3_QD-data}}
\end{figure}

\subsection{Quantum dot regime}

A majority of these devices show signatures of unintentional quantum dots formed inside the nanowire channel at low temperatures, due to random, static potential fluctuations. They are characterized by the observation of Coulomb diamonds in the $I-V$ characteristics at temperatures $T_\mathrm{L} \lesssim 10$ K. In most devices, the diamonds close indicating that a single QD dominates the transport. At $T_\mathrm{L} \lesssim 2$ K, superconducting correlations also appear in the nanowire channel due to its proximity to the Nb leads, interplaying with the QD charging energy and resulting in Andreev Bound States (ABS) associated with the QD. Here we focus on devices d3, d4, and d5 measured at $T_\mathrm{L} = 1.5$ K, representative of the total of $\sim 20$ measured devices.

Figure \ref{fig:3_QD-data}a, b, c shows numerical differential conductance $dI / dV$ measured at $T_\mathrm{L} = 1.5$ K for d3, d4, and d5, respectively, versus bias voltage $V$ and the back-gate voltage $V_g$. The data is characterized by the appearance of resonances at or below the Nb leads' superconducting gap at $\Delta \simeq 1.4$ meV, superimposed on the Coulomb diamond structure. We shall return to these resonant features shortly. Figure \ref{fig:3_QD-data}d shows the same data as panel c, but at a magnetic field $B = 3$ T perpendicular to the device substrate, which suppresses the superconductivity in the Nb leads. This allows the electron addition energy $E_{add}$ and the charge degeneracy points of the Coulomb diamonds to be measured. Lowering the field back down to $0$ T, we observe that any shift of the Coulomb diamond structure versus field is small, corresponding to a change in the backgate voltage $\delta V_g \lesssim 10$ mV. The QD charging energies (i.e. the smallest observed addition energies) are typically within the range $5 - 8$ meV, from which we estimate a QD radius on the order of $12 - 20$ nm. While based on a simple model of a spherical QD, this indicates that the QDs are small compared to the channel length $L = 170 - 200$ nm.

At zero magnetic field, peaks appear in conductance at resonant values of the bias voltage $V$, and these resonant values are modulated by changing the backgate voltage $V_g$. The dependence of the resonances on $V_g$ strongly correlates with the Coulomb diamond structure, as previously seen in carbon nanotube \cite{Kumar_PRB2014_CNT_ABS, Gaass_PRB2014_CNTABS, Pillet_PRB2013, Kim_PRL2013_CNTABS, Pillet_NatPhys2010, eichler_PRL2007_CNTABS} and InAs nanowire devices \cite{Jellinggaard_PRB2016, ChangW_nnano2015, ChangW_PRL2013, SandJespersen_PRL2007, DeFranceschi_nnano2014}. Unlike previous studies, however, there appears to be little correlation between the shape of the resonances versus $V_g$ and the charge state of the quantum dot; i.e. the even Coulomb valleys host similar resonances to the odd Coulomb valleys. This is in contrast with almost all previous observations in the literature, wherein the odd valleys host Yu-Shiba-Rusinov states that have a strong dependence on the gate voltage, but even valleys form resonances close to the superconducting gap $\Delta$, with little gate dependence (an exception is part of the data reported in reference \cite{Kumar_PRB2014_NbCntABS}). Below we use an extended Anderson-type model to qualitatively explain this behaviour. Another typical observation (see e.g. figure \ref{fig:3_QD-data}a) is for the resonances to be `pinned' to the bias voltage $|V| = \Delta$, or merge to form a zero-bias peak (ZBP) at $V = 0$; however, they can generally appear at any value $|V| < \Delta$. Importantly there are no features visible at $|V| = 2 \Delta$,  and contrary to observations of QDs formed in carbon nanotubes contacted with superconductors \cite{Pillet_NatPhys2010}, there is no transport `gap' in the low bias regime. These observations suggest that regions of the nanowire channel with finite (ungapped) density of states are connecting the QD to the Nb leads, i.e. an S-N-QD-N-S transport geometry, in which the QD is randomly placed inside the nanowire channel, with normal leads (N) connecting the QD on one or both sides to the superconducting (S) leads. The N-sections can carry superconducting correlations to the QD via the Andreev reflection process. The resonant features are identified as Andreev Bound States (ABS) associated with the QD, and a model for the ABS is presented below. Transport across the junction corresponds to energy resolved tunnelling of the ABS, with the normal sections of the nanowire corresponding to weakly coupled probes, and tunnelling occurring across the potential barriers at the edges of QD.

A good model for the ABS observed here must reproduce the following experimental observations: (i) for each value of $V_g$ there are two pairs of resonances, where each pair consists of a resonance at $ \pm V$. (ii) At the charge degeneracy points, the ABS at positive bias become degenerate at a bias value $ 0 < V < \Delta$, similarly for the two at negative bias, at $-\Delta < V < 0$. (iii) There is no discernible even-odd effect for the observed ZBP with respect to the electron number on the QD. This rules out Kondo correlations as a potential mechanism behind the ZBP: indeed, the Kondo temperature for the QDs is approximated, based on the Anderson model formula \cite{Haldane_PRL1978}, to be $T_{K} \simeq 2$ K, similar to previous reports on measurements of $T_K$ for InAs nanowire QDs \cite{Mahalu_2011_Kondo_InAs}. The experimental data is collected at an electron temperature of $1.5$ K $\sim T_{K}$, so it is not surprising if signatures of Kondo correlations cannot be observed. Preliminary data taken at $T_\mathrm{L} = 25$ mK does not show signatures of Kondo effects, but further experiments are required to rule it out completely. (iv) The formation (or lack thereof) of the ZBP appears to be correlated with the addition energy $E_{add}$ of the QD; for large $E_{add}$ the ABS appear to be pinned at $|V| = 0, \Delta$ (for gate voltages $V_g$ tuned away from the charge degeneracy points), whereas for small $E_{add}$ the ABS appears at $0< |V| < \Delta$ (see figure \ref{fig:3_QD-data}b,c).

\subsection{Model}

A slightly modified version of an Anderson-type model \cite{Anderson_PR1961, Pillet_PRB2013} detailed in Ref. \cite{kirsanskas_ysr_2015} is used to describe the observed ABS. The elements of the model are two superconducting (S) leads with energy gap $\Delta$, each tunnel coupled to a QD with a coupling strength $\Gamma_i$, where $i = L,R$ denote the left and right contacts. The superconducting phase difference between S-leads is denoted by $\phi$. The transport geometry within the model is an S-QD-S configuration. As mentioned earlier, we believe our devices to be in an S-N-QD-N-S configuration; however, neglecting the N-sections results in a much simplified model which can be used to accurately describe the experimental data, as described below. The device channel lengths $L = 170 - 200$ nm are assumed to be smaller than the phase coherence length of the channel at $1.5$ K, so the N-sections can transfer superconducting correlations to the QD through the Andreev reflection process. Within the context of the model, the presence of these N-sections have no effect on the physics of the ABS other than a rescaling of the S-QD coupling strengths $\Gamma_i$. In a full treatment of the problem, S-QD coupling will involve the details of the S-N tunnel coupling, the N-section transmission/scattering processes, and the N-QD tunnel coupling. Such considerations are left for future work, and the entire process is concisely described by the `effective' model parameters $\Gamma_i$. The cost of this simplification is that $\Gamma_i$ cannot be interpreted as tunnelling rates from which the total current can be derived, but rather as parameters describing the strength of S-QD correlations.

\subsubsection{Lack of an even-odd effect}

The Anderson model can describe Yu-Shiba-Rusinov (YSR) states arising in QDs connected to superconducting leads. The mechanism for YSR states relies on a spin impurity (in this case a spin-1/2) impurity inside the QD to couple to a spinful quasiparticle in the S lead. As such, YSR states are only predicted to exist for odd-occupation states of the QD in an S-QD-S system. For an even occupation of the QD, the ABS are predicted assume a simpler character, pinned to $\pm \Delta$ with little to no gate voltage modulation. Much experimental evidence points to this even-odd effect in the ABS spectrum of InAs, InSb and carbon nanotube S-QD-S devices. However, our observations, consistent across $\sim 20$ devices, are different. We observe YSR-like behaviour for both even- and odd-occupation Coulomb valleys of the QD. As mentioned above, only the addition energy $E_{add}$ appears to control the behaviour of the ABS resonances. 

The key to explaining this lack of even-odd behaviour is most likely related to the S-N-QD-N-S geometry of our devices and the nature of the N-sections. Disorder in the electrostatic potential is present in the whole nanowire, and in the N-sections it produces a `spiky' density of states (DOS) rather than the smooth DOS of an ideal one-dimensional conductor. This can also be viewed as weak charge localization, i.e. large quantum dots with small charging energy compared to $\Delta$ and weak tunneling barriers. Since the Andreev reflections between the Nb leads result in wavefunctions spanning the entire channel, any spin impurity within the channel can couple to the Nb leads to create YSR states \cite{Balatsky_RMP2006}. The presence of this non-trivial DOS in the N-sections affords the freedom to assume a spinful state inside the channel with little energy cost, regardless of the electron occupation of the primary QD. With this picture in mind, we apply analytical solutions for YSR states of an Anderson model to describe the ABS resonances in both the even- and odd-electron Coulomb valleys. 

\begin{figure}[t!]
	\centering
	\includegraphics[width=8.5cm]{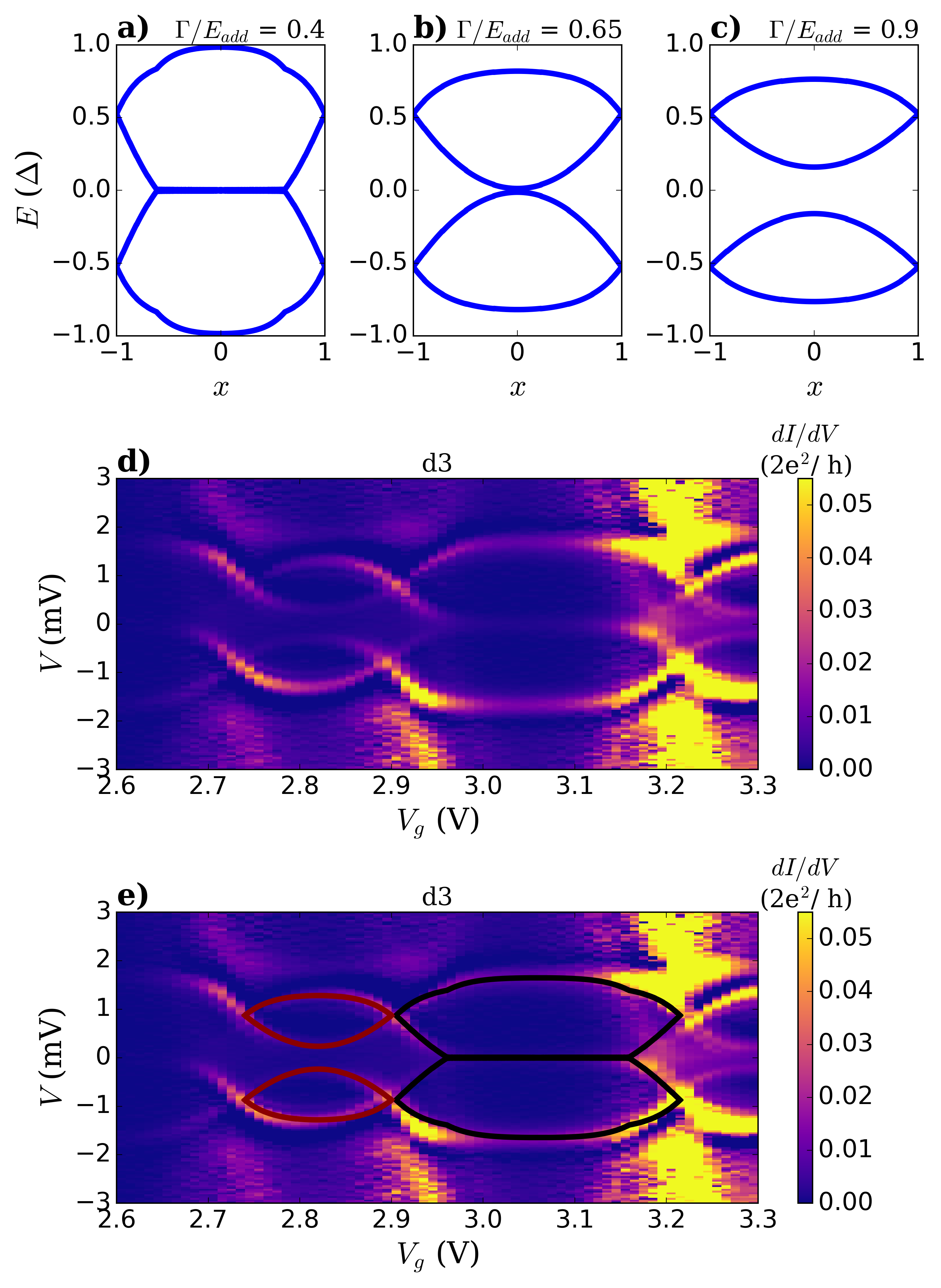}
	\caption{Modelled ABS energies (normalized to superconducting gap $\Delta$) vs normalized gate voltage $x$, as a function of $\Gamma / E_{add}$. For a large value of the electron addition energy $E_{add}$ ($\Gamma / E_{add} < 0.65$ in this example, panel \textbf{a}), two ABS are pinned at $|E| = \Delta$, and another ABS pair merge to form a ZBP over a large range of gate voltage $x$; whereas for a small value of $E_{add}$ ($\Gamma / E_{add} > 0.65$, panel \textbf{c}), four non-generate ABS occur at $0 <|E| < \Delta$ for all $x$, and panel \textbf{b} shows the intermediate case. \textbf{d)} Enlarged region of figure \ref{fig:3_QD-data}a. \textbf{e)} Same as \textbf{d}, with superimposed theoretical ABS resonances. The following parameters are used in generating the theoretical curves: $\Gamma / E_{add} = 0.87$ (cyan), $\Gamma / E_{add} = 0.43$ (dark blue), $\Delta = 1.4$ meV, $\Gamma_L / \Gamma_R = 0.57$, $\Gamma / \Delta = 4$, $T = 1.5$ K.
	  \label{fig:4_YSR-model}}
\end{figure}

\subsubsection{Subgap energies}

The subgap energies $E$ are calculated versus normalized gate voltage $x = 1 + 2\varepsilon/E_{add}$. Here, $\varepsilon$ is the chemical potential of the QD level, and $x = 1,-1$ refer to charge degeneracy points. The starting point is the analytical result obtained in Eq. 34 of Ref. \cite{kirsanskas_ysr_2015}. For each value of $x$, we choose a superconducting phase $\phi_0$ that will maximize the supercurrent $I_s (x,\phi)$ carried by the ABS, where $I_s(x,\phi)$ is calculated \cite{bagwell_prb1992, Schapers_2001_p72} using the formula
\begin{equation}
I_s(x,\phi) = \frac{2e}{h} \sum_n \left( f(E_n (x,\phi)) \times  \frac{\partial}{\partial \phi} E_n (x,\phi) \right).
\end{equation}
Here, $f$ is the Fermi-Dirac distribution at temperature $T = 1.5$ K, $e$ is the electronic charge, $h$ is Planck's constant, and $E_n$ is the energy level of the ABS as calculated in Eq. 34 of Ref. \cite{kirsanskas_ysr_2015}, where $n$ enumerates the four allowed solutions. Finding the optimal $\phi_0$ as a function of $x$ amounts to modelling the junction as \textit{current-biased}, instead of phase-biased. This is the appropriate choice for the S-N-QD-N-S geometry of devices studied here, because $\phi$ is not fixed by a magnetic flux threading a SQUID loop, but rather, the ABS levels are tunnel-probed by injecting current through them via the N-leads.

Figure \ref{fig:4_YSR-model}(a-c) shows the theoretical energies of the subgap ABS states versus $x$, and as a function of the $\Gamma/E_{add}$, where $\Gamma = \sqrt{\Gamma_L^2 + \Gamma_R^2}$ is the total effective S-QD coupling strength. It is seen that a ZBP forms due to a merger of two ABS for $\Gamma/E_{add} < 0.65$, starting at around $x=0$. Figure \ref{fig:4_YSR-model}d shows an enlarged version of figure \ref{fig:3_QD-data}a, and panel e shows the same data with the theoretical ABS curves superimposed. A coupling asymmetry value $\Gamma_L / \Gamma_R = 0.57$ is found to produce the best fit to the data. Two characteristic electron addition energies $E_{add}$ are seen in Figure \ref{fig:4_YSR-model}(d,e). The diamond with $E_{add} = 13$ meV is best fit to the model with $\Gamma / E_{add} = 0.43$ (dark blue), whereas the diamond with $E_{add} = 8$ meV is best fit to $\Gamma / E_{add} = 0.87$ (cyan). Data from d3 is shown as representative here; however, it is found that for all data sets, once model parameters $\Delta$ and the ratio $\Gamma_L / \Gamma_R$ are fixed, good agreement with the model can be reached by finding the optimal value for $\Gamma/E_{add}$ within each Coulomb diamond. We conclude that the simplified model presented here captures the basic physics of the current-biased ABS.

\section{\label{sec:discussion}Discussion}

\textit{Josephson regime} --  The interface transparency $t$ crucially affects the proximity gap induced inside the semiconductor by the superconducting leads. Several groups have recently focused on \textit{in-situ} grown, epitaxial Al contacts on InAs and InSb in order achieve $t \sim 1$. However, it was recently shown \cite{Zhang_Delft2016_MF} that a hard proximity gap can be induced in InSb nanowires using sputtered NbTiN contacts, with $t \sim 0.98$, proving that epitaxial contacts are not a strict requirement for high-quality interfaces. In order to optimize $t$ in our devices, current work is under way to improve the nanowire surface preparation process prior to the deposition of the Ti/Nb contacts, including passivation of the nanowire surface using a sulfur-rich ammonium sulfide solution. We have obtained preliminary results with S-passivated devices in which a Josephson supercurrent is observed even at $T_\mathrm{L} = 1.5$ K. While still in the diffusive regime, such devices are well suited to the search for signatures of Majorana fermions in the supercurrent \cite{Aguado_PRB2014_IcSignature}.\\

\textit{Decay of $I_c$ versus $B_{||}$} -- The decay of the critical current $I_c$ versus the axial magnetic field can be fit to a Gaussian curve up to $B_{||} \sim 0.8$ T for both d1 and d2 (figure \ref{fig:2_B-dep}), suggesting that a magnetic depairing mechanism \cite{Hammer2007} is at play. However, it is unclear why the depairing occurs $2 - 3$ times faster versus $B_{||}$ than predicted \cite{Hammer2007}. It is also unclear why the mechanism is not effective at suppressing the supercurrent for $B_{||} \gtrsim 0.8$ T until the upper critical field for the Nb contacts is reached at $H_{c2} \simeq 2.8$ T. An accurate description of the magnetic field depairing effect thus remains out of reach, likely requiring numerical solutions to the Usadel equations for our device geometry. Another possibility is that the initial decay is not dominated by the depairing effect, but rather a mesoscopic interference effect. For example, orbital Josephson interference \cite{Gharavi_PRB2015} can result in a sharp initial decay of $I_c$ versus $B_{||}$ if a large number of orbital angular momentum subbands are occupied.\\

\textit{Screening of magnetic flux by the Nb leads} -- Since the axial magnetic field $B_{||}$ is applied in the plane of the device substrate, the expected effect of the $50 - 80$ nm thick Nb leads is to screen (rather than focus) the magnetic flux by a small amount at the nanowire channel (see ref. \cite{Strambini_2016} for a similar effect). We estimate \footnote{Based on a magnetic penetration depth $\gtrsim 90$~nm for a $50 - 80$~nm thick Nb film \cite{Gubin_PRB2005_Nb_lambda}, and assuming an exponential suppression of the parallel field inside the leads.} that the value of the field at the center of the nanowire channel is a factor of $\lesssim 20 \%$ smaller than the applied field. Thus, the faster-than-expected decay of $I_c$ versus $B_{||}$ is not due to field focusing. \\

\textit{Quantum dot regime} -- ABS have previously been studied in InAs nanowires contacted with Al \cite{Jellinggaard_PRB2016, ChangW_nnano2015, ChangW_PRL2013, SandJespersen_PRL2007} and V \cite{DeFranceschi_nnano2014} and in carbon nanotube devices \cite{Kumar_PRB2014_CNT_ABS, Gaass_PRB2014_CNTABS, Pillet_PRB2013, Kim_PRL2013_CNTABS, Pillet_NatPhys2010, eichler_PRL2007_CNTABS}, as well as a range of other physical systems \cite{Janvier_sci2015, deon_APL2011, Buizert_PRL2007_aggABS}. To our knowledge, this report is the first description of ABS in quantum dots in InAs nanowires connected to Nb leads. The larger transition temperature and superconducting gap of Nb compared to Al allows the experiment to be performed at $1.5$ K, where the Kondo effect is not observed. A good agreement is reached between the Anderson-type model of ABS and the experimental data. This leads to exciting possibilities for further research on this system, including doublet-to-singlet transition of the ground state \cite{DeFranceschi_nnano2014, ChangW_PRL2013} and search for Majorana fermions in phase-biased ABS \cite{Hansen_PRB2016_N-SNS}.\\

\section{\label{sec:conclusions}Conclusions}
We have studied quantum transport in diffusive, short-channel InAs nanowire/Nb Josephson junctions, and identified two distinct transport regimes. Relatively large supercurrents and Nb/InAs contact transparencies in the Josephson regime are encouraging results, indicating that junctions based on InAs nanowires with non-epitaxial Nb contacts are good candidates for exploring proximity effect and Majorana physics at relatively high magnetic fields. The behavior of the critical current versus the axial magnetic field, including modulation and persistence to high field, is not yet understood. We presented several hypotheses to explain this, but further study is needed. These effects need to be understood so that they can be distinguished from signatures of a topological phase transition \cite{Albrecht2016, Zhang_Delft2016_MF, Mourik2012}. In the quantum dot regime, subgap resonances can be well described as current-biased Andreev Bound States within an Anderson-type model. To explain the lack of an even-odd effect in the data, we hypothesize that the N-section provides spinful states that induce YSR-like bound states associated with the quantum dot. Numerical calculations using the NRG technique could be used to further test this idea. The reproducibility of our experimental observations across $\sim 20$ devices compels further research on InAs nanowire/Nb non-ballistic junctions, especially in light of recent proposals \cite{Hansen_PRB2016_N-SNS, Huang_PRB_2014} for detecting signatures of Majorana fermions using phase-biased ABS.

\begin{ack}
We thank Karsten Flensbserg and Francois Sfigakis for helpful discussions. K.G. thanks Eduardo Barrera for technical assistance. Nanowire growth was performed at The Centre for Emerging Device Technologies at McMaster University. The University of Waterloo's Quantum NanoFab was used for device fabrication. This work was supported by the Natural Sciences and Engineering Research Council of Canada and the Ontario Ministry of Research \& Innovation. Quantum NanoFab acknowledges support from the Canada Foundation for Innovation, Industry Canada and Mike \& Ophelia Lazaridis.
\end{ack}

\section*{References}

\providecommand{\newblock}{}

\end{document}